\documentclass[conference]{IEEEtran}
\IEEEoverridecommandlockouts
\usepackage{cite}
\usepackage{amsmath,amssymb,amsfonts}
\usepackage{algorithm,algorithmic}
\usepackage{subfigure,graphicx}
\usepackage{textcomp}
\usepackage{xcolor}
\usepackage{multirow}
\usepackage{diagbox}
\usepackage{hyperref}
\hypersetup{
	colorlinks=true,
	linkcolor=blue,
	filecolor=blue,
	urlcolor=blue,
	citecolor=cyan,
}
\def\BibTeX{{\rm B\kern-.05em{\sc i\kern-.025em b}\kern-.08em
		T\kern-.1667em\lower.7ex\hbox{E}\kern-.125emX}}
\begin{document}
	\title{Communication under Mixed Gaussian-Impulsive Channel: An End-to-End Framework}
	\author{\IEEEauthorblockN{Chengjie Zhao, Jun Wang, Xiaonan Chen, Wei Huang, and Tianfu Qi}
	\IEEEauthorblockA{National Key Laboratory of Science and Technology on Communications \\
	University of Electronic Science and Technology of China, Chengdu, China\\
}
	}
	\maketitle
	
	\begin{abstract}
		In many communication scenarios, the communication signals simultaneously suffer from white Gaussian noise (WGN) and non-Gaussian impulsive noise (IN), i.e., mixed Gaussian-impulsive noise (MGIN). Under MGIN channel, classical communication signal schemes and corresponding detection methods usually can not achieve desirable performance as they are optimized with respect to WGN. Moreover, as the widely adopted IN model has no analytical and general closed-form expression of probability density function (PDF), it is extremely hard to obtain optimal communication signal and corresponding detection schemes based on classical stochastic signal processing theory. To circumvent these difficulties, we propose a data-driven end-to-end framework to address the communication signal design and detection under MGIN channel in this paper. In this proposed framework, a channel noise simulator (CNS) is elaborately designed based on an improved generative adversarial net (GAN) to simulate the MGIN without requirement of any analytical PDF. Meanwhile, a multi-level wavelet convolutional neural network (MWCNN) based preprocessing network is used to mitigate the negative effect of outliers due to the IN. Compared with conventional approaches and existing end-to-end systems, extensive simulation results verify that our proposed novel end-to-end communication system can achieve better performance in terms of bit-error rate (BER) under MGIN environments.
	\end{abstract}

	\section{Introduction}
	In communication systems, the noise is commonly modeled as Gaussian distribution in the light of the central limit theorem (CLT). Under this assumption, the optimal signal detection algorithms have been comprehensively investigated and perform well under a lot of circumstances. However, those detection algorithms would have poor performance while the transmitted signals are contaminated by non-Gaussian impulse noise \cite{nonlinearSP}. Impulse noise was confirmed to exist in many scenarios, such as low-frequency (LF) \cite{stanforddoc} and power-line \cite{Powerline} communication systems. Thus, it is well-motivated to address the communication issues under additive white Gaussian noise (AWGN) channel interfered by the impulse noise, i.e., mixed Gaussian and impulsive noise (MGIN) channels.
	
	Under non-Gaussian impulsive noise, the optimal and sub-optimal signal detection methods are commonly nonlinear \cite{nonlinearSP} and can be categorized into indirect detection \cite{awmyf, rwmyf} and direct detection \cite{P2BandJD, branchmeasuremsk}. For indirect detection methods,  nonlinear preprocessing is introduced to suppress the influence of impulsive noise on the transmitting signals. Then, the signal would be further detected by utilizing those techniques proposed with respect to Gaussian noise. A kind of widely used nonlinear preprocessing method is the Myriad filter (MyF) and its variants \cite{nonlinearSP}, \cite{awmyf, rwmyf}. On the other hand, direct detection is based on the maximum likelihood (ML) criterion derived from the probability density function (PDF) of noise instead of preprocessing. As the PDF of non-Gaussian noise in practice is often extremely complicated and usually not closed, characteristic function or approximate PDF has been proposed to design detection methods \cite{mixednoise2}. However, the aforementioned approaches under non-Gaussian noise are not optimal due to the inherent discrepancy between practical MGIN and the adopted models \cite{mixednoise1}. Correspondingly, the resulting solution suffers from a significant performance gap compared to the optimum.
	
	In the last decade, deep learning (DL) has achieved rapid development and has been successfully applied in many fields due to its outstanding capability in fitting sophisticated models. DL has also been employed in the physical layer of communication systems and shown great potential, such as DL based signal detection algorithm \cite{DLbasedSD}, model-driven block-wise DL optimization schemes \cite{modeldriven}, end-to-end global optimization communication systems \cite{e2eTOS, E2ELY}.
	
	Inspired by the aforementioned works, it could be expected that DL can be adopted in communication systems under the MGIN channel to obtain better performance compared to the traditional methods. It is noteworthy that \cite{THe2e} recently considered DL for communication under non-Gaussian noise. However,  the achieved performance is not desirable. Therefore, we further propose a DL-based end-to-end communication framework under MGIN by fully considering the characteristics and influence of impulsive noise in this paper. In our framework, a generative adversarial net (GAN) \cite{GAN} is adopted to model the MGIN distribution and used as the channel noise simulator (CNS) of the proposed end-to-end system. To overcome the inherent mode collapse problem of GAN, we adopt the method presented in \cite{WGANGP} and introduce a special regularization term to constrain the training process. Moreover, we apply a preprocessing network to tackle the outliers due to impulse noise. The preprocessor is based on a multi-level wavelet convolutional neural network (MWCNN) \cite{MWCNN} framework which has been applied for image restoration under Gaussian noise. Consequently, the proposed framework can work well in various noise environments.

	Extensive numerical simulations are performed to verify the advantages of the proposed communication system. Compared with existing DL-based and conventional methods, simulation results show that our proposed communication framework can achieve better performance in terms of bit-error rate (BER) and well adapt to various MGIN channels.
	
	\section{End-to-End Framework}
\begin{figure}[htbp]
		\centering
		\includegraphics[scale=0.6]{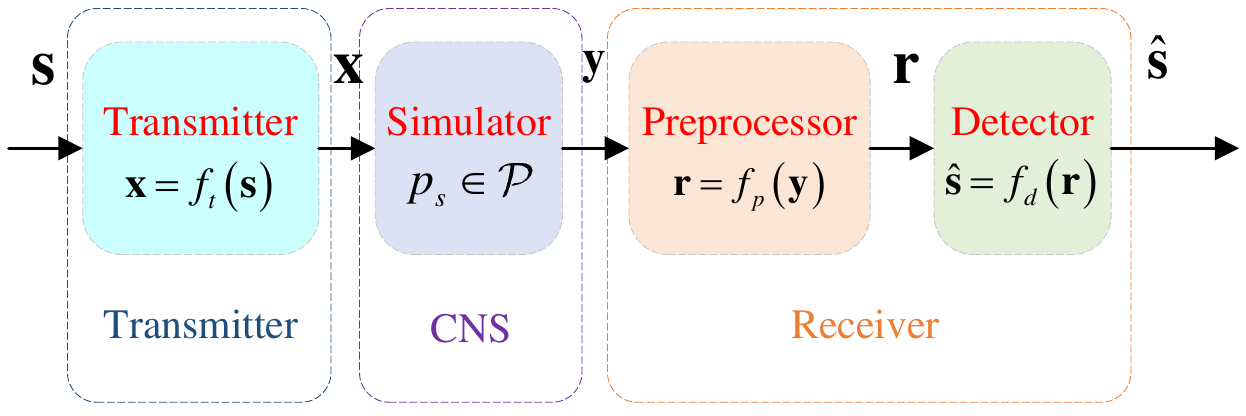}
		\caption{The architecture of proposed end-to-end system}	
		\label{fig1}
	\end{figure}

	The architecture of our proposed end-to-end framework is shown in Fig.\ref{fig1}. It is composed of four parts, including the transmitter, CNS, preprocessor, and detector. All the components are implemented via neural networks (NN). From a DL perspective, the whole system could be regarded as an Auto-Encoder architecture in which the transmitter and receiver take the role of encoder and decoder, respectively.
	
	At the transmitter, the independent and identically distributed (i.i.d.) information bits sequence with length $L$ is denoted as ${\bf{s}} = \left[ {{s_1},{s_2}, \cdots ,{s_L}} \right]$ and ${s_i} \in \left\{ {0,1} \right\},i \in \left\{ {1,2, \cdots L} \right\}$. The goal of the transmitter is mapping $\textbf{s}$ to generate transmitting signal sequence ${\mathbf{x}} = \left[ {{x_1},{x_2}, \cdots ,{x_N}} \right] \in {\mathbb{R}^N}$ that well matches the MGIN channel. Here, $N$ denotes the transmitting signal sequence length, and $L/N\leq 1$ can be considered as the system code rate. As an encoder, the transmitter tries to effectively represent the input information based on the channel characteristics. This task can be performed by learning an optimal mapping function $\mathbf{x}={f_t}\left(\textbf{s} \right)$ via an NN.
	
	The CNS is employed to fit the MGIN distribution without any analytical and closed-form PDF expressions. Then, the original process that the transmitting signals are influenced by MGIN can be equivalent to passing through the CNS. This process could be written as
	\begin{equation}
		{\mathbf{y}} = {\mathbf{x}} + {\mathbf{m}} = {\mathbf{x}} + {{\mathbf{m}}_g} + {{\mathbf{m}}_s} \Leftrightarrow {\mathbf{y}} = {\mathbf{x}} + {\mathbf{g}}
	\end{equation}
	where $\mathbf{g} \in {\mathbb{R}^N}$ denotes the noise vector generated by the CNS, ${\mathbf{m}}_g$ and ${\mathbf{m}}_s$ are the Gaussian and impulsive noise components of MGIN with length $N$, respectively. Usually, ${\mathbf{m}}_g$ and ${\mathbf{m}}_s$ can be considered to be independent.
	
	Both the preprocessor and detector constitute the receiver. The receiver aims to recover the transmitting information bit sequence ${\mathbf{s}}$ from the distorted signal $\mathbf{y} \in {\mathbb{R}^N}$. Specifically, the preprocessor is applied to combat with the outliers in $\mathbf{y}$ produced by impulsive noise. The resulting $\mathbf{r} \in {\mathbb{R}^N}$ is fed into the detector to obtain the estimation ${\mathbf{\hat s}}$ of $\mathbf{s}$.
	
	\section{Channel Noise Simulator}
	The channel noise simulator aims to capture the characteristics of MGIN by fitting the PDF of the MGIN. In this paper, we adopt the GAN architecture by taking advantage of its flexibility in defining objective functions. In what follows, we focus on the objective function design of GAN to meet the requirement of simulating MGIN.
	
\begin{figure}[htbp]
		\centering
		\includegraphics[scale=0.55]{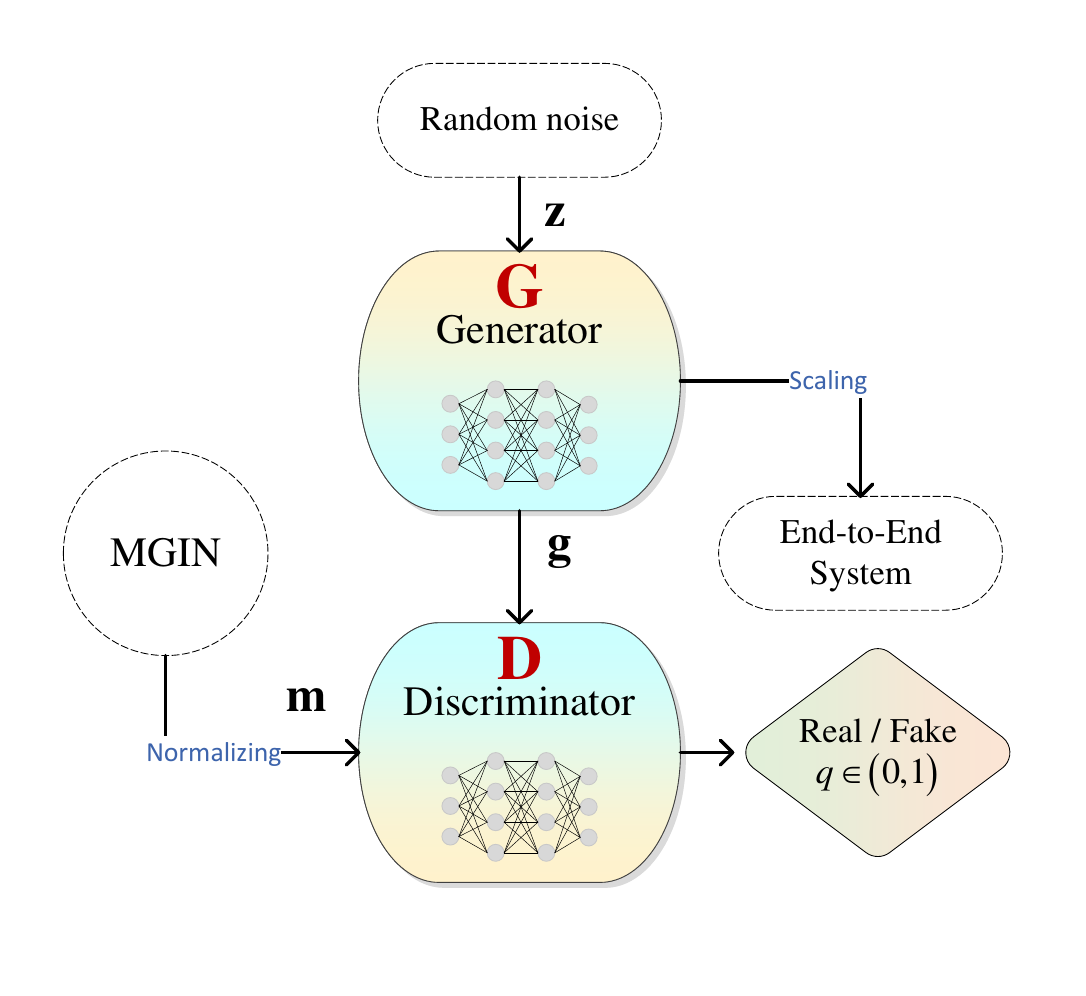}
		\caption{The structure of GAN for MGIN simulation}	
		\label{fig2}
	\end{figure}

	GAN is an adversarial generative architecture that consists of two counterparts, including the generator $G$ and the discriminator $D$, as shown in Fig.\ref{fig2}. $G$ learns to simulate the distribution of MGIN, while $D$ is responsible for identifying whether a sequence comes from the environment or $G$. When $D$ cannot distinguish the output of $G$ from MGIN, the GAN reaches its optimum.

	As illustrated in Fig. \ref{fig2}, a random noise sequence $\bf{z}$ with a prior distribution $p_z$ is fed into generator $G$ to generate a noise sequence $\mathbf{g}$ with distribution $p_e$, which is expected to be as close as the distribution of MGIN $p_m$. $p_z$ usually adopts Gaussian distribution. Generator $G$ represents a mapping $G\left( {{\bf{z}};{\theta _g}} \right)$ with parameter $\theta_g$, which transfer noise $\bf{z}$ into a data space distributed as $p_g$.  Discriminator $D$ is another mapping $D\left( {{\bf{d}};{\theta _d}} \right)$ with parameter $\theta_d$ and its input ${\bf{d}}$, that gives a scalar output $q \in \left( {0,1} \right)$ to indicate the probability that a sequence comes from the MGIN rather than $G$. The ${\bf{d}}$ could be ${\bf{g}}$ or ${\bf{m}}$.
	
	The goal of training the adopted GAN is to make the output data space of $G$ as close to MGIN noise space as possible, and $D$ could correctly identify the noise sample $\bf{m}$ and generated sample $\bf{g}$. Jensen-Shannon (JS) divergence is usually used to formulate the objective function in training a GAN \cite{GAN}. As the original GAN is unstable in training, Wasserstein distance (WD) and gradient penalty are introduced to improve the training in \cite{WGANGP}. The objective functions could be written as
	\begin{equation}\label{Eq_LD}
		\begin{aligned}
			{L_D}\left( {{\theta _d}} \right) &= {L_{\mathbf{z}}}\left( {{\theta _d}} \right) - {L_{\mathbf{d}}}\left( {{\theta _d}} \right) + \lambda \operatorname{gp} \left( {{\theta _d}} \right) \\
			&= {E_{{\mathbf{z}} \sim {p_z}\left( {\mathbf{z}} \right)}}\left[ {D\left( {\mathbf{g}} \right)} \right] - {E_{{\mathbf{m}} \sim {p_m}\left( {\mathbf{m}} \right)}}\left[ {D\left( {\mathbf{m}} \right)} \right] \\
			& \qquad + \lambda {E_{{\mathbf{\hat m}} \sim {p_{\hat m}}\left( {{\mathbf{\hat m}}} \right)}}\left[ {{{\left( {{{\left\| {{\nabla _{{\mathbf{\hat m}}}}D\left( {{\mathbf{\hat m}}} \right)} \right\|}_2} - 1} \right)}^2}} \right] \\
		\end{aligned}
	\end{equation}
	\begin{equation}\label{Eq_LG}
		{L_G}\left( {{\theta _g}} \right) =  - {L_{\mathbf{z}}}\left( {{\theta _g}} \right) = - {E_{{\mathbf{z}} \sim {p_z}\left( {\mathbf{z}} \right)}}\left[ {D\left( {\mathbf{g}} \right)} \right]
	\end{equation}
	where $\lambda$ denotes penalty coefficient, $\operatorname{gp}(\cdot)$ is the gradient penalty term that ensures the 1-Lipschitz constraint. Sample $\mathbf{\hat m}$ is a uniformly sampled convex combination of MGIN $\mathbf{m}$ and generated $\mathbf{g}$, and $p_{\hat m}$ represents its distribution. ${\nabla _x}f\left( x \right)$ denotes the gradient of $f$ with respect to $x$. However, the inherent mode collapse still exists when GAN is applied to simulate MGIN, which has complicated distribution. Thus, we further propose several training strategies to well fit MGIN distribution.
	
	Before proceeding, we first introduce two functions. The first one has a U shape. We call it the U function, which can be expressed as follows
	\begin{equation}\label{U_Function}
		{\operatorname{U} _{{x_m}}}\left( x \right) = \frac{1}{{\sqrt {1 - {{\left( {\frac{x}{{{x_m}}}} \right)}^2}} }},\left| x \right| < {x_m},{x_m} > 0
	\end{equation}
	where $x_m$ is the boundary. Note that U function together with zero-order Bessel function of the first kind $\pi {x_m}{J_0}\left( {2\pi {x_m}t} \right)$ are Fourier transform pair. The most significant characteristic of U function \eqref{U_Function} is that the large value is close to the boundary. It can be utilized to generate outliers by combining them with corresponding training skills. Another function is $\operatorname{V} \left( {{\mathbf{a}},{\mathbf{b}}} \right) = {\left( {{{\left\| {\mathbf{a}} \right\|}_2} - {{\left\| {\mathbf{b}} \right\|}_2}} \right)^2}$, which can indicate the length difference between two vectors or the energy variance between two signals. This function could help to shape the PDF of generated data. In the training of GAN, these two functions could be used as the constraints for impulsive noise and Gaussian noise components, respectively.
	
	The training strategy is also illustrated in Fig.\ref{fig2}. The proposed GAN obtains MGIN sequences from the environment and then normalize them to the range of $\left[ { - 1,1} \right]$, so that $G$ could utilize a \textit{Tanh} activation layer at its output end to enhance its nonlinear capacity. The \textit{Tanh} activation function is denoted as $\operatorname{Tanh} \left( x \right) = \frac{{\sinh \left( x \right)}}{{\cosh \left( x \right)}} = \frac{{{e^x} - {e^{ - x}}}}{{{e^x} + {e^{ - x}}}}$. Simultaneously, the boundary of U function regularization could take a value $1 + \epsilon$. Here, $\epsilon < {\mathcal{O}}(1)$ is a constant that guarantees such a range is in the domain of U function regularization. The values close to 1 would receive a reward so that the generator would pay attention to the outliers and consequently overcome the mode collapse. Hence, \eqref{Eq_LG} could be reformulated as
	\begin{equation}\label{Eq_AdoptedLG}
		\begin{aligned}
			{L_G}\left( {{\theta _g}} \right) &=  - {L_{\mathbf{z}}}\left( {{\theta _g}} \right) - \frac{\alpha_1}{N} \cdot {\operatorname{U} _{1 + \epsilon }}\left( {\mathbf{g}} \right) + \frac{\alpha_2}{N} \cdot {\mathbf{V}}\left( {{\mathbf{g}},{\mathbf{m}}} \right) \hfill \\
			&=  - {E_{{\mathbf{z}} \sim {p_z}\left( {\mathbf{z}} \right)}}\left[ {D\left( {\mathbf{g}} \right)} \right] - \frac{\alpha_1}{N} \cdot \sum\limits_{n = 1}^N {\frac{1}{{\sqrt {1 - {{\left( {\frac{{{g_n}}}{{1 + \epsilon }}} \right)}^2}} }}}  \hfill \\
			& \qquad + \frac{\alpha_2}{N} \cdot {\left( {{{\left\| {\mathbf{g}} \right\|}_2} - {{\left\| {\mathbf{m}} \right\|}_2}} \right)^2} \hfill \\
		\end{aligned}
	\end{equation}
	where $\alpha_1$ and $\alpha_2$ respectively denote the weight coefficient of two regularization terms, and $x_m$ is replaced by $1 + \epsilon$. The $\epsilon$, $\alpha _1$, and $\alpha _2$ are all tunable parameters that can be adjusted to adapt to various MGIN environments.

	\section{Preprocessor at Receiver}
	Under MGIN channel, the outliers introduced by the impulsive noise component will significantly deteriorate the communication waveform and lead to severe performance degradation. Usually, it is rather difficult to suppress this kind of outliers as well as avoid damaging the communication waveform. To avert the difficulty, we adopt an NN-based preprocessor to suppress the outliers.
	
\begin{figure}[htbp]
		\centering
		\includegraphics[scale=0.25]{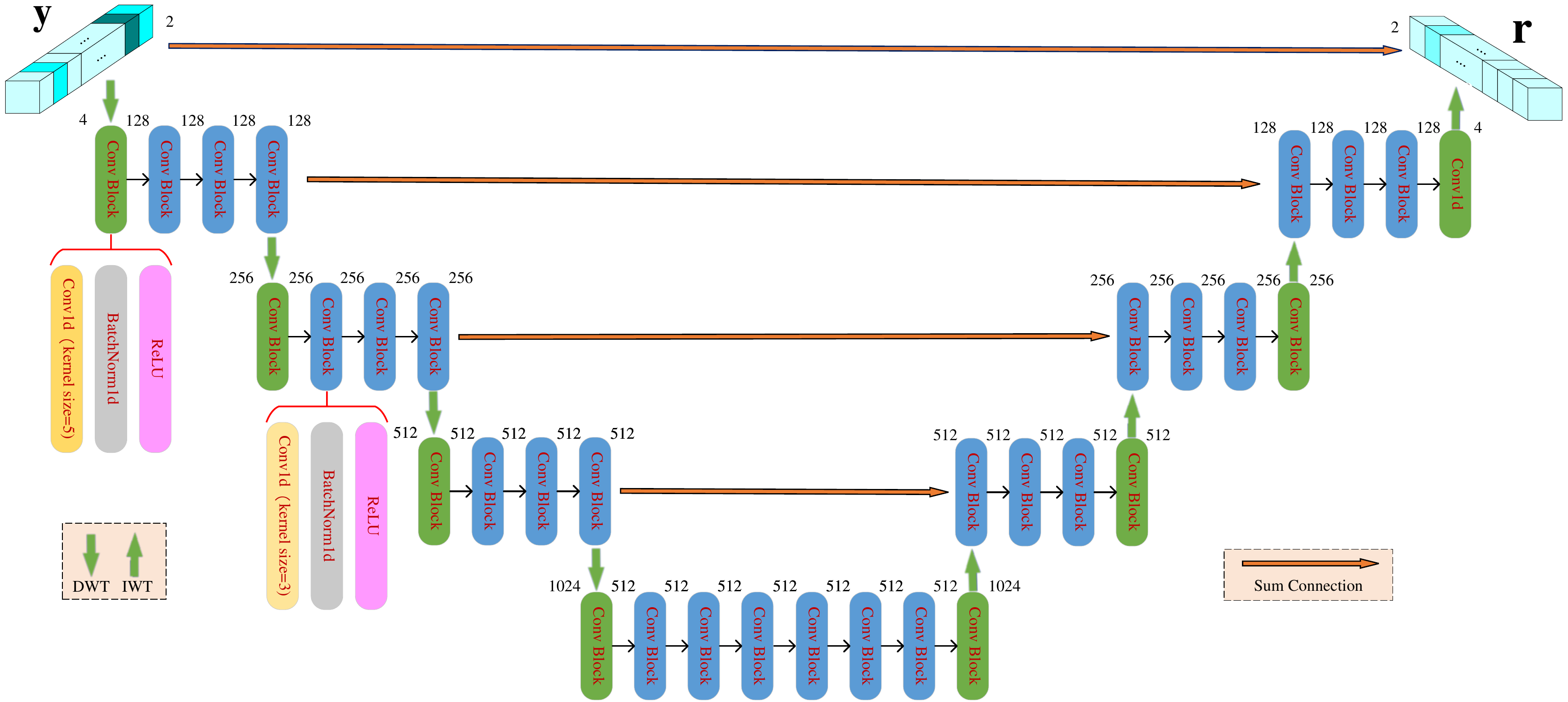}
		\caption{The structure of Sequence MWCNN}
		\label{fig3}
	\end{figure}

	For the preprocessor, it should be robust to adapt to various MGIN channels and preserve the information of interest as much as possible. Meanwhile, a proper trade-off between processing capability and complexity is also needed. To achieve these goals, we adopt an MWCNN \cite{MWCNN} to act as the preprocessor in this paper. The MWCNN has been proved to be effective in the field of image denoising and achieves a good compromise between capability and computational cost \cite{MWCNN}. The MWCNN also possesses an elegant U-shaped architecture, that excels at capturing input information. Meanwhile, instead of the pooling layers, the discrete wavelet transformation (DWT) are applied to guarantee the exploitation of input information. Different from image processing, this preprocessing is a 1-dimensional signal processing problem. Therefore, we call after this MWCNN as sequence MWCNN. The structure of our proposed MWCNN is shown in Fig.\ref{fig3}.

	The essence of MWCNN is the stacked layers of DWT or inverse wavelet transformation (IWT) and the following CNN blocks. For MWCNN, the widely used pooling layers and up-convolution layers are replaced by DWT/IWT so that the loss of information during the processing procedure can be avoided.
	
	At each level of the downstream decomposition process, the contaminated signal is first performed DWT to separate its high-frequency and low-frequency components. Then, the separated components pass through a cascade of CNN blocks to compact their representations. Similar to the downstream, a series of CNN blocks rearrange the decomposed signals, and the resulting signals are then expanded by IWT in the upstream recovery process. Simultaneously, the summation connections are applied to bridge the feature maps in the equal level of downstream and upstream.
	
	The adopted MWCNN is jointly trained with the other components. The overall training procedure will be described in the next subsection.
	
	\section{Training Overall Transceiver}
	
	As a communication system, the objective of the system is to minimize the end-to-end bit error. The objective function could be expressed as
	\begin{equation}
		L\left( {{\theta _t},{\theta _p},{\theta _{de}}} \right) = \operatorname{d} \left( {{\mathbf{s}},{\mathbf{\hat s}}} \right)
	\end{equation}
	where $\theta _t$, $\theta _p$, and $\theta _{de}$ represent the parameters of the transmitter, preprocessor, and detector, respectively. $\operatorname{d} \left( {{\mathbf{x}},{\mathbf{y}}} \right)$ denotes a measure of the difference between two vectors $\mathbf{x}$ and $\mathbf{y}$. We considered two practicable metrics to measure the difference. One is the binary cross-entropy loss, and the other is the mean square error loss. The corresponding objective function could be reformulated as
	\begin{equation}\label{Eq_Lb}
		{L_b}\left( {{\theta _t},{\theta _p},{\theta _{de}}} \right) = \frac{1}{N}\sum\limits_{i = 1}^N {{s_i}\log \left( {{{\hat s}_i}} \right) + \left( {1 - {s_i}} \right)\log \left( {1 - {{\hat s}_i}} \right)}
	\end{equation}
	\begin{equation}\label{Eq_Lm}
		{L_m}\left( {{\theta _t},{\theta _p},{\theta _{de}}} \right) = \frac{1}{N}\sum\limits_{i = 1}^N {{{\left( {{s_i} - {{\hat s}_i}} \right)}^2}}
	\end{equation}
	respectively. By taking these two metrics into consideration, the task of our system could be deemed as classification or waveform recovery, respectively.
	
\begin{figure}[htp!]
		\centering
		\includegraphics[scale=0.65]{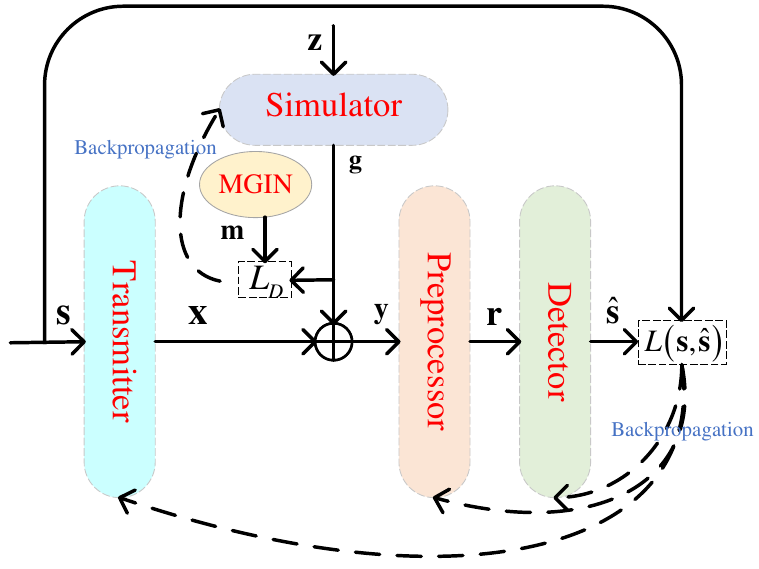}
		\caption{The training procedure of whole system}
		\label{fig4}
	\end{figure}

	The transmitter, preprocessor, and detector can be alternatively and iteratively updated under the supervision of the end-to-end loss \eqref{Eq_Lb} or \eqref{Eq_Lm}. \textit{BatchNorm1d} layers are introduced to avoid the problem of vanishing gradient in such a DNN. The training procedure is shown in Fig.\ref{fig4}.
	
	Besides, the CNS can be trained jointly with the above three components, or be separately trained at first. Anyway, the learned simulator could provide output highly similar to MGIN to serve the training of the entire system.
	
	\section{Experiment Results}
	In this section, we verify the performance advantage of our proposed end-to-end communication system through simulations.
	
	\subsection{MGIN Settings}
	As mentioned before, the MGIN could be modeled as $m = {m_g} + {m_s}$, where $m_g \sim \mathcal{N}\left(0,\sigma ^2\right)$ is a Gaussian random variable with variance $\sigma ^2$, and $m_s \sim S_\alpha{\left(\gamma,0,0\right)}$ is a S$\alpha$S random variable with dispersion $\gamma$ and characteristic exponent $\alpha$ \cite{sassummarize}. In fact, Gaussian distribution $\mathcal{N}\left(0,\sigma ^2\right)$ corresponds to S$\alpha$S distribution ${S_2}\left( {\frac{{{\sigma ^2}}}{2},0,0} \right)$. We control the intensity of impulsive noise in MGIN by adjusting the ratio $\lambda = 2\gamma:\sigma^2$. The stronger the impulsive noise is, the larger the $\lambda$ is. We would perform the simulation with various $\lambda$ to show the adaptability of our system.
	
	We adopt generalized ${{{E_b}} \mathord{\left/{\vphantom {{{E_b}} {{N_0}}}} \right. \kern-\nulldelimiterspace} {{N_0}}}$ to represent the strength of signal against the MGIN. As impulsive noise has no finite second-order moment, we define the generalized signal-to-noise ratio (GSNR) as
	$$\text{GSNR} = \frac{{{P_s}}}{{{\sigma ^2} + 2\gamma }}$$
	where $P_s$ denotes the average power of the signal. The relationship between GSNR and ${{{E_b}} \mathord{\left/{\vphantom {{{E_b}} {{N_0}}}} \right. \kern-\nulldelimiterspace} {{N_0}}}$ is
	$${{{E_b}} \mathord{\left/{\vphantom {{{E_b}} {{N_0}}}} \right. \kern-\nulldelimiterspace} {{N_0}}} = \text{GSNR} \cdot \frac{{{T_b}}}{{{T_s}}} / {\log _2}M$$
	where $T_s$, $T_b$, and $M$ represent the sampling interval, symbol period, and modulation order, respectively. For a two-order modulation scheme, we have
	$${{{E_b}} \mathord{\left/ {\vphantom {{{E_b}} {{N_0}}}} \right. \kern-\nulldelimiterspace} {{N_0}}}\left( \text{dB} \right) = \text{GSNR}\left( \text{dB} \right) + 10{\log _{10}}\frac{{{T_b}}}{{{T_s}}}.$$
	\subsection {Adopted NN models}
		
	In simulations, the adopted network models for MWCNN have already illustrated in Fig.\ref{fig3}. The utilized network models and corresponding parameters for the GAN, i.e., CNS, transmitter and detector are provided in Table \ref{tab1} and \ref{tab2}, respectively. Meanwhile, batch size is set to 128.
	\begin{table}[htpb!]
		\centering
		\caption{Model Parameters of Proposed GAN}
		\label{tab1}
		\begin{tabular}{|cc|}
			\hline
			\multicolumn{1}{|c|}{\textbf{Type of Layer}}    & \textbf{Parameters}    \\ \hline
			\multicolumn{2}{|c|}{\textbf{Generator $\theta_g$}}                                 \\ \hline
			\multicolumn{1}{|c|}{Linear + BN1d + LeakyReLU} & (100,128), slope=0.2   \\ \hline
			\multicolumn{1}{|c|}{Linear + BN1d + LeakyReLU} & (128,256), slope=0.2   \\ \hline
			\multicolumn{1}{|c|}{Linear + BN1d + LeakyReLU} & (256,512), slope=0.2   \\ \hline
			\multicolumn{1}{|c|}{Linear + BN1d + LeakyReLU} & (512,256), slope=0.2   \\ \hline
			\multicolumn{1}{|c|}{Linear + Tanh}             & (256,256)              \\ \hline
			\multicolumn{2}{|c|}{\textbf{Discriminator $\theta_d$}}                             \\ \hline
			\multicolumn{1}{|c|}{Linear + LeakyReLU}        & (256,128), slope=0.2   \\ \hline
			\multicolumn{1}{|c|}{Linear + LeakyReLU}        & (128,64), slope=0.2    \\ \hline
			\multicolumn{1}{|c|}{Linear}                    & (64,1)                 \\ \hline
			\multicolumn{1}{|c|}{\textbf{Optimizer}}        & \textbf{Learning Rate} \\ \hline
			\multicolumn{1}{|c|}{Adam}                   & 0.00005                \\ \hline
		\end{tabular}
	\end{table}
	\begin{table}[!htbp]
		\centering
		\caption{Model Structure of Transmitter and Detector}
		\label{tab2}
		\begin{tabular}{|ccc|}
			\hline
			\multicolumn{1}{|c|}{\textbf{Type of Layer}} & \multicolumn{1}{c|}{\textbf{Feature Numbers}} & \textbf{Kernel Size} \\ \hline
			\multicolumn{3}{|c|}{\textbf{Transmitter $\theta_t$}}                                                                          \\ \hline
			\multicolumn{1}{|c|}{Conv1d + BN1d + ReLU}   & \multicolumn{1}{c|}{(1,128)}                  & 5                    \\ \hline
			\multicolumn{1}{|c|}{Conv1d + BN1d + ReLU}   & \multicolumn{1}{c|}{(128,256)}                & 3                    \\ \hline
			\multicolumn{1}{|c|}{Conv1d + BN1d + ReLU}   & \multicolumn{1}{c|}{(256,64)}                 & 3                    \\ \hline
			\multicolumn{1}{|c|}{Conv1d + Tanh}          & \multicolumn{1}{c|}{(64,2)}                   & 3                    \\ \hline
			\multicolumn{3}{|c|}{\textbf{Detector $\theta_{de}$}}                                                                             \\ \hline
			\multicolumn{1}{|c|}{Conv1d + BN1d + ReLU}   & \multicolumn{1}{c|}{(2,64)}                   & 5                    \\ \hline
			\multicolumn{1}{|c|}{Conv1d + BN1d + ReLU}   & \multicolumn{1}{c|}{(64,128)}                 & 3                    \\ \hline
			\multicolumn{1}{|c|}{Conv1d + BN1d + ReLU}   & \multicolumn{1}{c|}{(128,256)}                & 3                    \\ \hline
			\multicolumn{1}{|c|}{Conv1d + BN1d + ReLU}   & \multicolumn{1}{c|}{(256,128)}                & 3                    \\ \hline
			\multicolumn{1}{|c|}{Conv1d + BN1d + ReLU}   & \multicolumn{1}{c|}{(128,64)}                 & 3                    \\ \hline
			\multicolumn{1}{|c|}{Conv1d + Sigmoid}       & \multicolumn{1}{c|}{(64,1)}                   & 3                    \\ \hline
			\multicolumn{1}{|c|}{\textbf{Optimizer}}     & \multicolumn{2}{c|}{\textbf{Learning Rate}}                          \\ \hline
			\multicolumn{1}{|c|}{Adam}                   & \multicolumn{2}{c|}{0.0005}                                         \\ \hline
		\end{tabular}
	\end{table}

	\subsection{MGIN Fitting Effect}
	As there is no general and closed-form PDF for most of MGIN, \cite{mixednoise2} provides an alternative approximate PDF. In order to measure the effectiveness of different approaches, we adopt the WD to indicate the distance between the original MGIN and the corresponding approximations. WD, also known as the earth-mover distance of distribution $P$ and $Q$, is defined as \cite{WD}
	$$\operatorname{W} \left( {P,Q} \right) = \mathop {\inf }\limits_{\pi  \in \Gamma \left( {P,Q} \right)} \int_{\mathbb{R} \times \mathbb{R}} {\left| {p - q} \right|d\pi \left( {p,q} \right)}$$
	where ${\Gamma \left( {P,Q} \right)}$ is the set of joint distributions on $\mathbb{R} \times \mathbb{R}$ whose marginals are $P$ and $Q$, respectively. The WD could be considered as the minimum cost that transforms one distribution into another. It is noteworthy that the WD could act on two distributions with different support sets, and is more sensible than other measurements, e.g., JS divergence.
	
\begin{figure}[!htbp]
		\centering
		\includegraphics[scale=0.32]{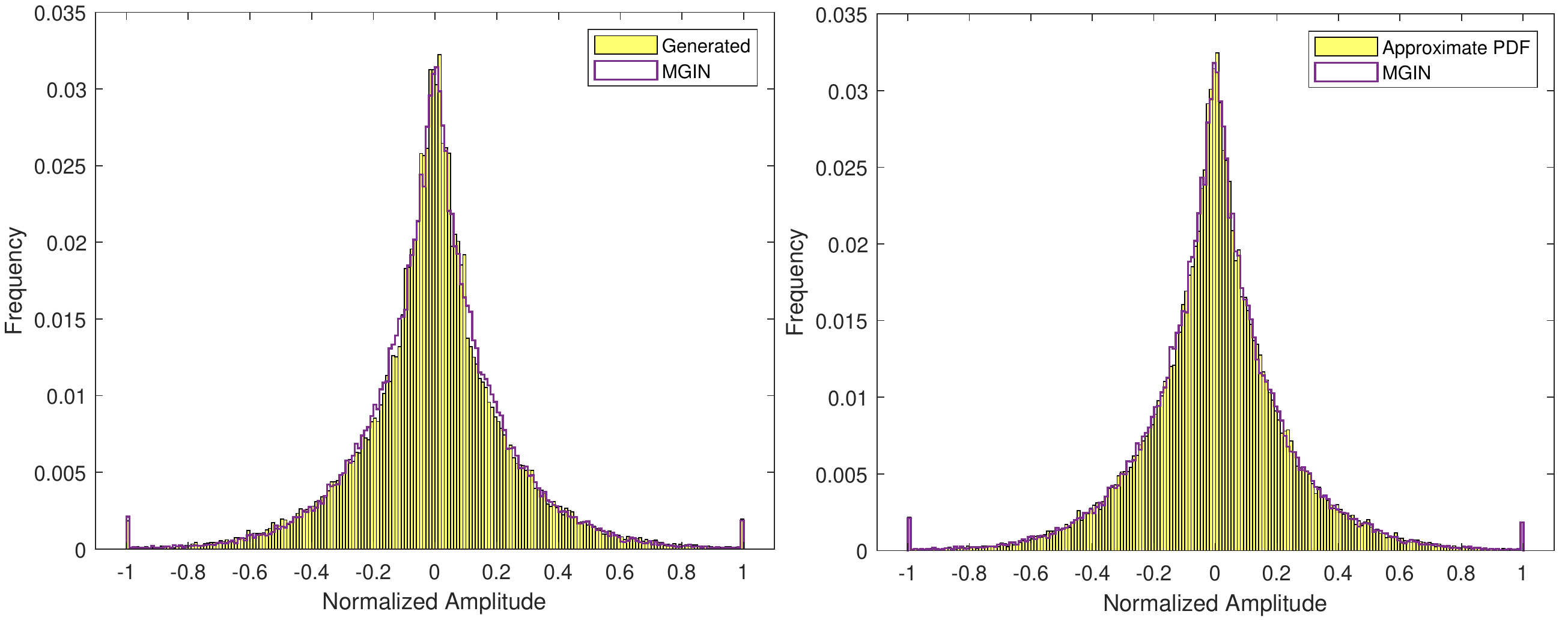}
		\caption{The histogram comparison under $\alpha=1.5, \lambda=\frac{1}{10}$}
		\label{fig5}
	\end{figure}

	We consider three scenarios, including the stronger Gaussian noise component, equal Gaussian noise and impulsive noise components, and stronger impulsive noise component. The corresponding $\lambda$ equals to $\frac{1}{10}, 1$, and $10$ respectively. We compare WD between the simulated PDF of our proposed GAN and the approximate PDF given in \cite{mixednoise2} with respect to the original MGIN under various circumstances. The comparison is shown in Table \ref{tab3}. In Fig. \ref{fig5}, we also provide statistical histograms to illustrate the fitting effect of our GAN and approximate PDF under the scenario $\alpha=1.5$ and $\lambda=1/10$.
	\begin{table}[h]
		\centering
		\caption{WD comparison}
		\label{tab3}
		\begin{tabular}{|cc|c|c|}
			\hline
			\multicolumn{2}{|l|}{\diagbox{MGIN}{WD}{Approach}}                   & GAN & Approximate PDF \\ \hline
			\multicolumn{1}{|c|}{\multirow{3}{*}{$\alpha=1.5$}} & $\lambda=\frac{1}{10}$ &  0.011   &  \textbf{0.009}               \\ \cline{2-4}
			\multicolumn{1}{|c|}{}                       & $\lambda=1$   &  \textbf{0.007}  &     0.008          \\ \cline{2-4}
			\multicolumn{1}{|c|}{}                       & $\lambda=10$   &   \textbf{0.008}  &     0.012            \\ \hline
		\end{tabular}
	\end{table}
	
	It can be found that our proposed GAN can simulate MGIN rather well, and achieve a similar fitting performance as that of approximate PDF. Note that cumbersome parameters estimation is needed to get the approximate PDF \cite{mixednoise2}.
	
	\subsection{End-to-End Performance}
	We further test the performance in terms of BER of our proposed end-to-end communication system under several typical MGIN environments. We would compare our system performance with some baseline approaches.
	
	\begin{figure}[!h]
		\centering
		\includegraphics[scale=0.5]{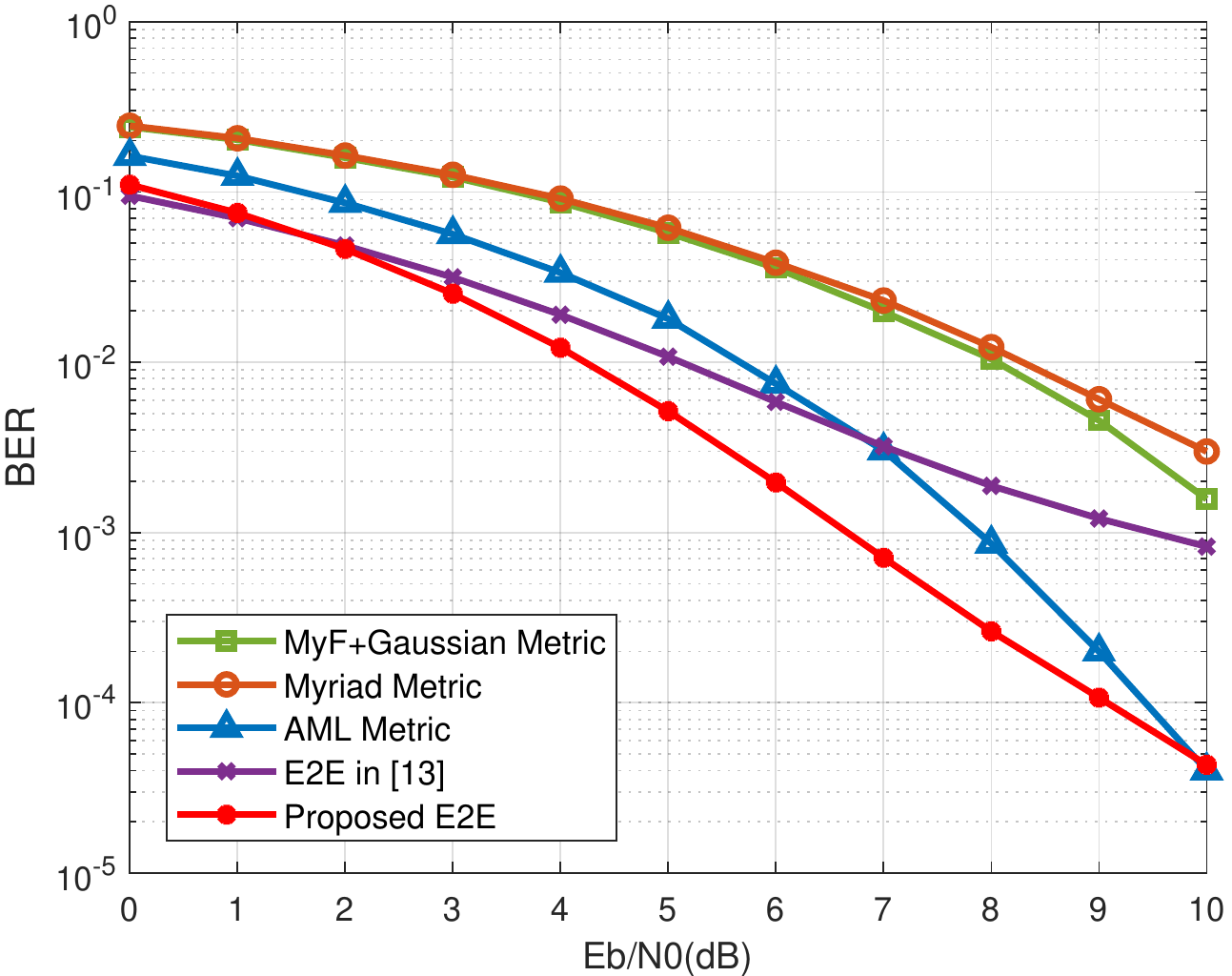}
		\caption{BER comparison under $\alpha=1.5, \lambda=\frac{1}{10}$}
		\label{fig6}
	\end{figure}
	\begin{figure}[!h]
		\centering
		\includegraphics[scale=0.5]{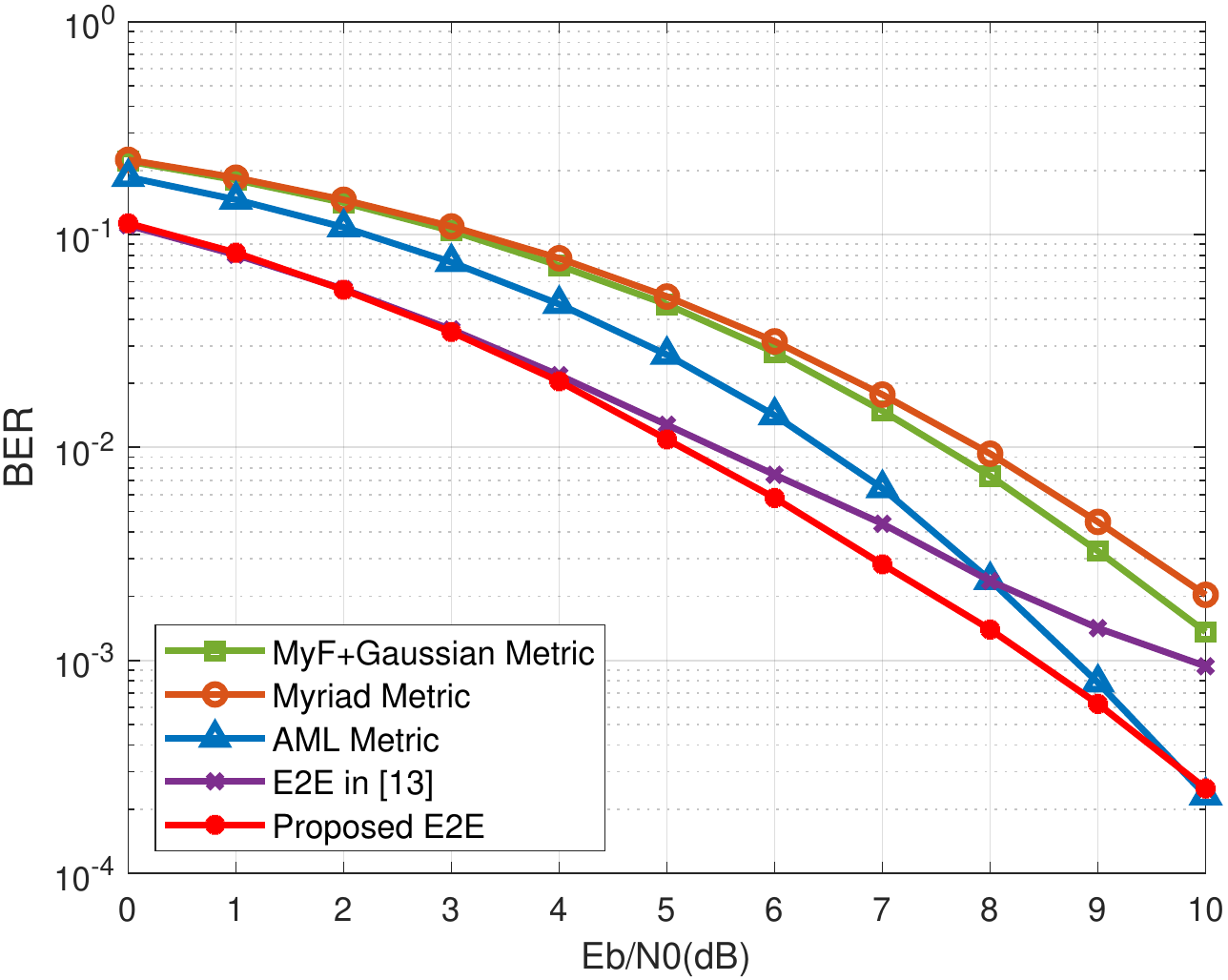}
		\caption{BER comparison under $\alpha=1.5, \lambda=1$}
		\label{fig7}
	\end{figure}
	\begin{figure}[!h]
		\centering
		\includegraphics[scale=0.5]{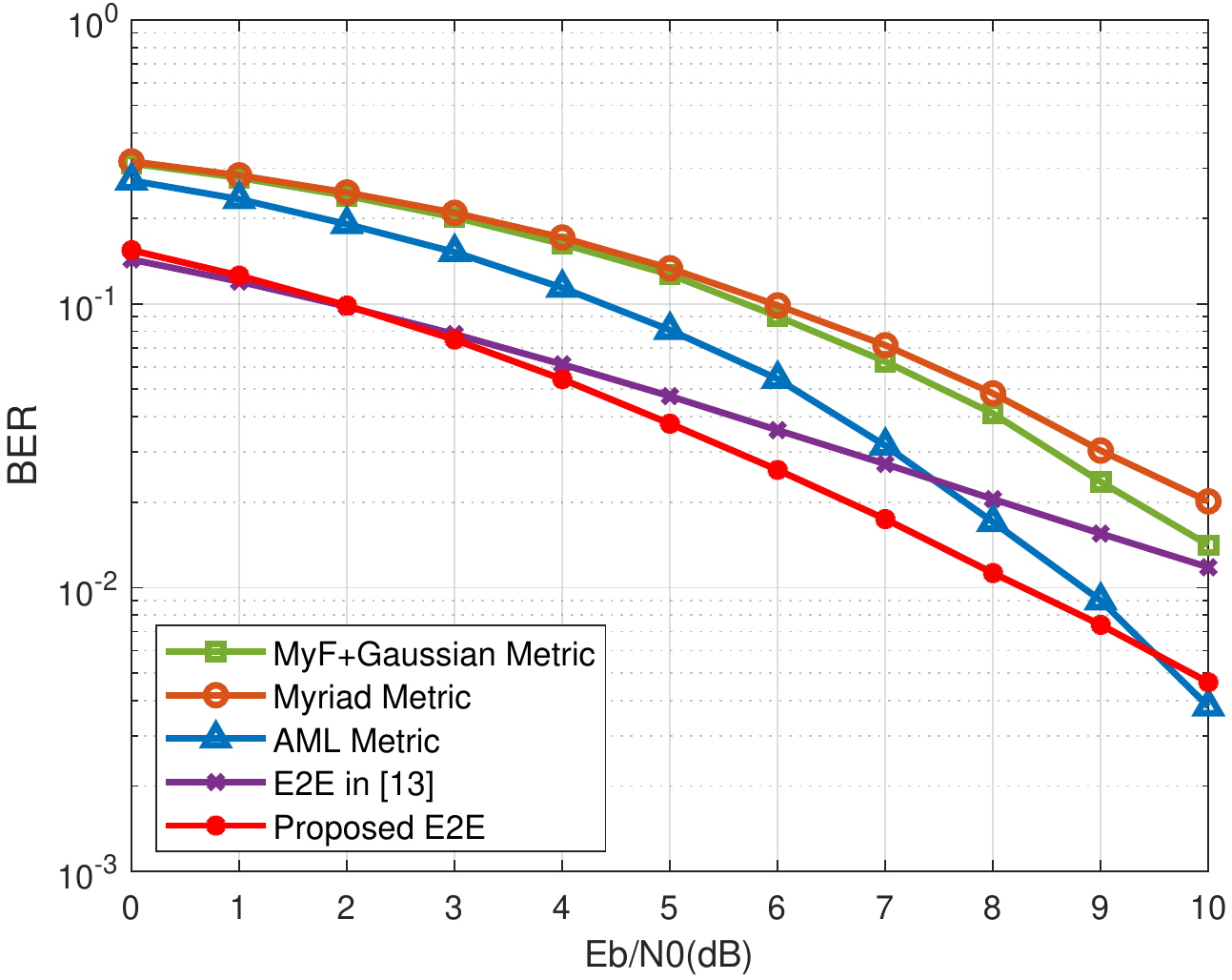}
		\caption{BER comparison under $\alpha=1.5, \lambda=10$}
		\label{fig8}
	\end{figure}
	
	Firstly, we consider the additive MGIN channel and test our proposed system under above mentioned three different impulse intensity. For conventional system, we focus on the minimum-shift-keying (MSK) modulation. MSK is widely used in the VLF/LF communication systems which suffer from impulsive noise. The Viterbi Algorithm (VA)-based sequence detection is used to detect the MSK signals \cite{branchmeasuremsk}. The Myriad, approximate ML (AML) metrics based on approximate PDF of MGIN, and Gaussian metric after MyF preprocessing are adopted for VA. Meanwhile, the DL-based end-to-end communication system presented in \cite{E2ELY} is selected as baselines. Both our proposed system and the system presented in \cite{E2ELY} are trained under ${{{E_b}} \mathord{\left/ {\vphantom {{{E_b}} {{N_0}}}} \right. \kern-\nulldelimiterspace} {{N_0}}} = 10{\text{dB}}$.
	
	In Fig.\ref{fig6}-\ref{fig8}, we present the BER performance comparison between our end-to-end system and the other baselines under different additive MGIN channels. It could be observed that the AML Metric-based method performs best among the conventional methods. The Myriad Metric-based method as well as the MyF preprocessing plus Gaussian Metric-based method reach the similar performance. Both of them are inferior to AML Metric-based method. Regardless of the intensity of impulsive noise, our proposed system could remarkably outperform the best AML Metric-based method when the signal strength is relatively weak and achieve similar BER for the case of a strong signal. Meanwhile, the existing end-to-end system is significantly inferior to our proposed end-to-end system under the MGIN scenario, as it is not elaborately optimized with respect to MGIN.

	 \begin{figure}[htbp]
	 	\centering
	 	\includegraphics[scale=0.5]{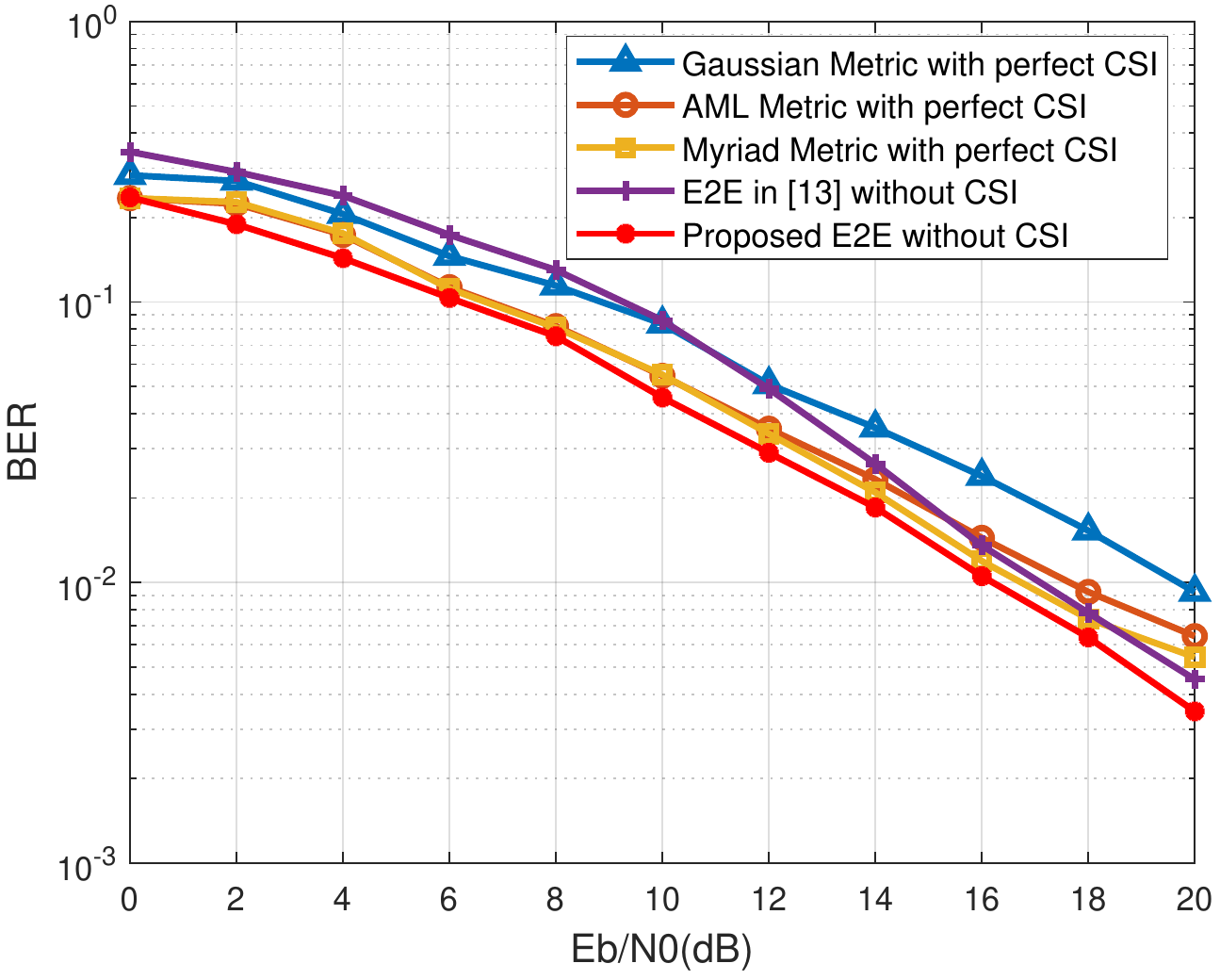}
	 	\caption{BER comparison under fading channel}
	 	\label{fig9}
	 \end{figure}	

	Furthermore, we consider Rayleigh fading channel with MGIN with $\alpha=1.5, \lambda=1$. We assume the conventional MSK approaches possess perfect channel state information (CSI), and zero-forcing (ZF) equalization is used to compensate fading. Meanwhile, the end-to-end systems process the signals directly without CSI, and are trained under ${{{E_b}} \mathord{\left/ {\vphantom {{{E_b}} {{N_0}}}} \right. \kern-\nulldelimiterspace} {{N_0}}} = 20{\text{dB}}$. As shown in Fig.\ref{fig9}, among the conventional methods, the Myriad and AML metric can merely perform a little better than the Gaussian metric. The performance of original specially optimized AML metric is even inferior than that of Myriad metric under some $\vphantom {{{E_b}} {{N_0}}}$ conditions for the original distribution of MGIN is distorted after the ZF equalization. On the other hand, our proposed end-to-end system could still achieve the best performance.
	
	\section{Conclusion and Discussion}
	In this paper, we introduce DL to tackle the communication signal design and detection under MGIN via an end-to-end framework. By considering the characteristics of MGIN, we artfully design NN architectures for the transmitter, CNS, and receiver. Compared with conventional approaches and an existing end-to-end system, extensive simulation results validate that the proposed DL-based end-to-end communication system has remarkable advantage in terms of BER performance.
	
	\bibliographystyle{IEEEtran}
	\bibliography{Communication_under_Mixed_Gaussian-Impulsive_Channel_An_End-to-End_Framework}
\end{document}